\documentclass[prb,amsmath,amssymb,twocolumn,superscriptaddress,floatfix]{revtex4}
\usepackage{bbm}
\usepackage{graphicx}
\usepackage{dcolumn}
\usepackage{bm}
\usepackage{subfigure}
\usepackage{amsmath}
\usepackage{feynmf}
\usepackage{hyperref}

\usepackage{attachfile}
\newcommand{\ua}{\uparrow}
\newcommand{\da}{\downarrow}
\newcommand{\ra}{\rightarrow}

\newcommand{\bs}{\boldsymbol}

\newcommand{\SRO}{Sr$_2$RuO$_4$}

\usepackage{times}

\begin{document}
\title{Exotic Cooper pairing in multi-orbital models of \SRO}
\author{Wen Huang}
\affiliation{Shenzhen Institute for Quantum Science and Engineering and Department of Physics, Southern University of Science and Technology, Shenzhen 518055, China}
\affiliation{Institute for Advanced Study, Tsinghua University, Beijing 100084, China}
\affiliation{Center for Quantum Computing, Peng Cheng Laboratory, Shenzhen 518005, China}
\author{Yi Zhou}
\affiliation{Beijing National Laboratory for Condensed Matter Physics $\&$ Institute of Physics, Chinese Academy of Sciences, Beijing 100190, China}
\affiliation{Kavli Institute for Theoretical Sciences $\&$ CAS Center for Excellence in Topological Quantum Computation, University of Chinese Academy of Sciences, Beijing 100190, China}
\author{Hong Yao}
\affiliation{Institute for Advanced Study, Tsinghua University, Beijing 100084, China}
\affiliation{ State Key Laboratory of Low Dimensional Quantum Physics, Tsinghua University, Beijing 100084, China }
\date{\today}

\begin{abstract}
The unconventional superconductivity in \SRO~continues to defy a unified interpretation. In this paper, we focus on some novel aspects of its superconducting pairing by exploiting the orbital degree of freedom in this material. The multi-orbital nature, combined with the symmetry of the orbitals involved, leads to a plethora of exotic Cooper pairings not accessible in single-orbital systems. Essential physics is illustrated first using a two-orbital model with $d_{xz}$- and $d_{yz}$-orbitals. We classify the gap functions according to the underlying lattice symmetries, analyze the effective theories of a few representative pairings, and make connections to \SRO~in the course. In particular, we show how spin-orbit coupling may entangle spin-triplet and spin-singlet pairings. For completeness, the classification is generalized to the three-orbital model involving the $d_{xy}$-orbital as well. The orbital-basis approach distinguishes from the itinerant-band description for \SRO, and hence offers an alternative perspective to investigate its enigmatic superconducting state. 
\end{abstract}

\maketitle
\section{Introduction}
Superconductivity in \SRO~was discovered a quarter of a century ago \cite{Maeno:94}. Widely hailed as an archetypal unconventional superconductor, no consensus is yet available regarding its pairing symmetry \cite{Rice:95,Mackenzie:03,Kallin:09,Kallin:12,Maeno:12,Liu:15,Kallin:16,Mackenzie:17}. Indications of spin-triplet \cite{Ishida:98,Duffy:00}, odd-parity \cite{Nelson:04} pairing with spontaneous time-reversal symmetry breaking (TRSB) \cite{Luke:98,Xia:06} were reported in a series of earlier measurements. Taken together, they point to a chiral $p$-wave order with d-vector $(k_x + ik_y)\hat{z}$, which may exhibit nontrivial topology and host exotic excitations such as Majorana zero modes. Such a pairing is also supported by a number of other measurements \cite{Kallin:09,Kallin:12,Maeno:12,Mackenzie:17}. However, this interpretation stands at odds with a variety of signatures not easily reconcilable with this chiral $p$-wave pairing \cite{Kallin:09,Kallin:12,Maeno:12,Mackenzie:17}, including the indications of nodal excitations \cite{Nishizaki:00,Hassinger:17}, the absence of spontaneous surface current \cite{Kirtley:07,Hicks:10,Curran:14} and the anomalous behavior under in-plane magnetic fields \cite{Deguchi:02,Yonezawa:14,Kuhn:17} and in-plane uniaxial strains \cite{Hicks:14,Steppke:17,Watson:18}. The out-of-plane d-vector orientation is further challenged by a recent observation of a Knight shift drop below $T_c$ under in-plane magnetic fields \cite{Luo:19}. Thus far, we still lack a pairing state that is able to coherently interpret all of the key experiments. It is hence sensible to both examine the existing theories and assumptions, and to search for alternative superconducting pairings that may ultimately bring a unified understanding.

\SRO~has three Fermi sheets derived mainly from the Ru $4d$ $t_{2g}$-orbitals \cite{Damascelli:00,Bergemann:00}. As superconductivity appears to emerge out of a coherent Fermi liquid \cite{Bergemann:03}, plenty of microscopic theories take an itinerant-electron perspective, in which only intra-band superconducting pairing is active although multiple bands are considered \cite{Nomura:00,Nomura:02,Raghu:10b,Huo:13,Wang:13,Scaffidi:14,Tsuchiizu:15,Huang:16,Zhang:17b}. In this setting, only electrons near the Fermi level are considered relevant to Cooper pairing. The resultant superconductivity, in one way or another, is driven by spin or charge fluctuations reminiscent of the celebrated Kohn-Luttinger mechanism \cite{Kohn:65}. The gap classification in the corresponding band basis is relatively straightforward \cite{Sigrist:91}. In the presence of finite SOC, spins are no longer good quantum numbers. Nonetheless, an effective pseudospin basis can be adopted \cite{Scaffidi:14,Zhang:17b}, thanks to the conservation of the Kramers degeneracy in the Bloch bands. An alternative approach is the orbital-basis description. In this description, Cooper pairs are formed by electrons with well-defined orbital characters \cite{Puetter:12}. Although a corresponding full-fledged symmetry classification is lacking, many existing studies on the phenomenology of the superconducting \SRO~are constructed on the multi-orbital basis (e.g. some recent studies in Refs. \onlinecite{Taylor:12,Annett:12,Chung:12,Ramires:16,Komendova:17}). When transformed into band basis, the state typically allows for interband pairing, which is crucial for the appearance of the intrinsic anomalous Hall effect (which leads to Kerr rotation \cite{Xia:06}) below $T_c$ in a multiband chiral p-wave superconductor \cite{Taylor:12,Annett:12,Wang:17}.

\begin{figure}[b]
\includegraphics[width=4.5cm]{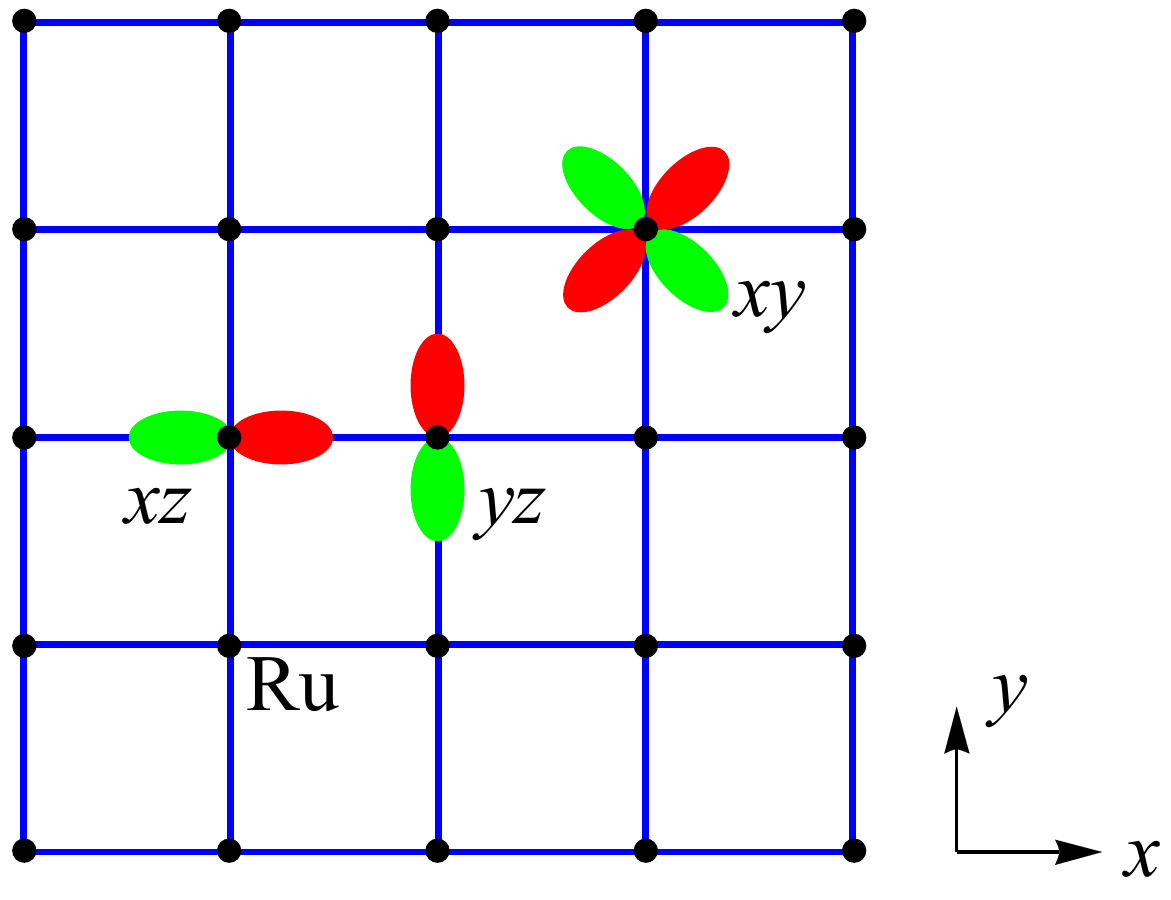}
\caption{Top view of  the $xz$-, $yz$-, and $xy$-orbitals on a 2D square lattice.  }
\label{fig:lattice}
\end{figure}

There are without doubt marked distinction between the band- and orbital-basis approaches. As we shall see in the present study, the latter exhibits a rich variety of exotic superconducting pairings. We illustrate this using a toy two-orbital model with the $t_{2g}$ $d_{xz}$- and $d_{yz}$-orbitals. Similar to some of the previous studies on the multi-orbital iron-based superconductors \cite{Dai:08,Zhou:08,Wan:08,Vafek:17,Cheung:19}, the gap functions are classified according to the underlying lattice point group symmetry. The orbital manifold in \SRO~introduces numerous novel possibilities not available in single-orbital models, such as even-parity spin-triplet and odd-parity spin-singlet pairings. We analyze the phenomenology of these states and discuss their possible relation to \SRO~when necessary. In particular, we show explicitly the influence of SOC on mixing spin-singlet and spin-triplet pairings in various superconducting channels \cite{Veenstra:14}. For completeness, we also present a gap classification for the three-orbital model that takes into account the $d_{xy}$-orbital as well.

\section{Single-particle Hamiltonian and gap classification}
\label{sec:singleH}
To make connection with \SRO, we take a two-orbital model with $d_{xz}$ and $d_{yz}$ orbitals residing on the each site of a square lattice (see Fig.\ref{fig:lattice} for illustration). The model contains no sublattice degree of freedom. In two spatial dimensions (2d), the model also applies to systems of $p_x$ and $p_y$ orbitals. It is instructive to first construct a continuum model Hamiltonian that respects both time-reversal and the $D_{4h}$ point group symmetries. In the spinor basis $(c_{xz\ua},c_{xz\da},c_{yz\ua},c_{yz\da})^T$,
\begin{equation}
H_{0\bs k}= t(k_x^2+k_y^2)-\mu+ \tilde{t}(k_x^2-k_y^2)\sigma_z + t^{\prime\prime}k_xk_y\sigma_x + \eta\sigma_y \otimes s_z  \,,
\label{eq:H0}
\end{equation}
where $\sigma_i$ and $s_i$ with $i=x,y,z$ are the Pauli matrices operating on the respective orbital and spin degrees of freedom, $(t,\tilde{t},t^{\prime\prime})$ designate the kinetic energy and $\eta$ the onsite spin-orbit coupling (SOC). This Hamiltonian is manifestly invariant under time-reversal, $\mathcal{T} = \sigma_0\otimes is_y \mathcal{K}$, where $\mathcal{K}$ denotes complex conjugation. It is also consistent with the tight-binding construction in previous studies \cite{}.

To see how Eq. \ref{eq:H0} respects $D_{4h}$, it is important to recognize that the point group operations must act jointly on spatial, spin and orbital degrees of freedom. This involves varying the phase (gauge) of the orbital wavefunctions under certain operations, due to the peculiar symmetry properties of the two orbitals. For example, a $C_4$ rotation, in addition to rotating momentum and spin, also exchanges the label of the two orbitals and induces a $\pi$ phase change on one of them, e.g. $(d_{xz},d_{yz}) \ra (d_{yz},-d_{xz})$. As a consequence, the bilinear $\sigma$-operators, which are formally $c^\dagger_{m,s}\sigma_i^{mn} c_{n,s^\prime}$ $(m=xz,yz)$, transform according to irreps of $D_{4h}$ in the following fashion \cite{Wan:08}: $\sigma_0$, $\sigma_x$, $\sigma_z$ and $\sigma_y$ as $A_{1g}$, $B_{2g}$, $B_{1g}$ and $A_{2g}$, respectively. A Hamiltonian invariant under all $D_{4h}$ operations is then constructed by appropriate product of the $\sigma$, $s$, and momentum-space basis functions, as in Eq (\ref{eq:H0}). Note that, amongst the terms in the Hamiltonian, $s_z$ transforms as $A_{2g}$. Further, since the orbital wavefunction of the $t_{2g}$-electrons are even under inversion, the only effect of inversion is to invert electron momentum. This differs from the model with $p_x$ and $p_y$-orbitals, where inversion also changes the sign of the fermion creation and annihilation operators (the bilinear operators are however unaffected by this). Taken together, it can be verified that the Hamiltonian Eq. (\ref{eq:H0}) respects the full $D_{4h}$ symmetry.

The possible pairing symmetries are typically classified according to the irreducible representations (irreps) of the underlying crystalline point group. This is straightforwad in single-orbital models, as have been well documented by Sigrist and Ueda \cite{Sigrist:91}. However, the presence of multiple orbitals adds a layer of complexity. The usual classifications into even-parity spin-singlet and odd-parity spin-triplet pairings are no longer sufficient. One must also consider Cooper pairs symmetric and anti-symmetric in the orbital manifold \cite{Zhou:08,Wan:08,Vafek:17,Cheung:19}. In addition, care must be taken of the transformation properties of the bilinear pairing operators $c_{m,s}\sigma_i^{mn}c_{n,s^\prime}$, analogous to that of $c^\dagger_{m,s}\sigma_i^{mn}c_{n,s^\prime}$ mentioned above.

Tables \ref{tab:tab1} lists the representative superconducting basis functions in different irreps of the $D_{4h}$ group. We see that most of the individual irreps contain multiple symmetry-equivalent basis functions -- a prominent feature not present in single-orbital systems. Note that, the spatial parity is a good quantum number in this system, and basis functions even and odd in ${\bs k}$ will not mix, because the inversion operation only acts to invert the momentum ${\bs k}$ while leaves orbital and spin degrees of freedom unchanged. This is quite different from systems with sublattice degree of freedom, such as a honeycomb lattice, where the gap functions may comprise components even and odd in momentum (overall inversion symmetry is nonetheless retained).

In the following, we develop effective Ginzburg-Landau theories for a few representative irreps. The first example, the $A_{1g}$ irrep, serves to illustrate how the multiple symmetry-equivalent multi-orbital pairings in each irrep may be inherently coupled by inter-orbital hybridization and SOC. In particular, this analysis will make transparent the SOC-induced entanglement of spin-triplet and spin-singlet pairings. We then proceed to the two-dimensional $E_g$ and $E_u$ irreps in view of the signatures of two-component superconducting pairing in \SRO. 

\begin{table*}[t]
\caption{\label{tab:table1}  Representative basis functions of the superconducting pairing in the two-orbital model in various irreps of the $D_{4h}$ point group. Here $\sigma_i$ and $s_i$ operate in the orbital and spin space, respectively. The vectors ${\bs x}$, ${\bs y}$ and ${\bs z}$ denote the direction of the d-vector of spin-triplet pairings, and the pairing gap functions are obtained by multiplying the basis function by $is_y$ (same below). Throughout this work, we neglect out-of-plane pairing for simplicity (see Sec \ref{sec:summary}).}
{\renewcommand{\arraystretch}{1.3}
\begin{tabular}{c|c}
\hline
~~~~irrep.~~~~ & ~~~~~~~~~basis function ~~~~~~~\\
\hline
$A_{1g}$   & $i\sigma_y\otimes \bs z\!\cdot\! \bs s $,~~$\bs 1$,~~$k_xk_y \sigma_x $,~~$(k_x^2-k_y^2)\sigma_z$ \\
\hline

$A_{2g}$   &  $k_xk_y\sigma_z $,~~$(k_x^2-k_y^2)\sigma_x $  \\
\hline

$B_{1g}$   &  $\sigma_z $,~~$i\sigma_y\otimes (k_x^2-k_y^2)\bs z \!\cdot\! \bs s $  \\
\hline

$B_{2g}$   &  $\sigma_x $,~~ $i\sigma_y\otimes k_xk_y\bs z \!\cdot\! \bs s $  \\
\hline

\shortstack{ ~~ \\ $E_g$ \\ ~~ \\ ~~\\~~}             &

\shortstack{~~ \\ $\left( i\sigma_y\otimes  \bs x \!\cdot\! \bs s ,~~i\sigma_y\otimes \bs y \!\cdot\! \bs s  \right)$ \\
~~\\
$\left[i\sigma_y\otimes  k^2_{x(y)}  \bs x \!\cdot\! \bs s,~~i\sigma_y\otimes k^2_{y(x)}  \bs y \!\cdot\! \bs s  \right] $ }\\

\hline

$A_{1u}$   & {$\frac{\sigma_0\pm \sigma_z}{2}\otimes k_x\bs x \!\cdot\! \bs s+ \frac{\sigma_0\mp \sigma_z}{2}\otimes k_y\bs y \!\cdot\! \bs s $,~~ $\sigma_x\otimes (k_x \bs y + k_y \bs x)\!\cdot\! \bs s $}  \\
\hline

$A_{2u}$   & $\frac{\sigma_0\pm \sigma_z}{2}\otimes k_x\bs y \!\cdot\! \bs s- \frac{\sigma_0\mp \sigma_z}{2}\otimes k_y\bs x \!\cdot\! \bs s $,~~ $\sigma_x\otimes (k_x \bs x - k_y \bs y)\!\cdot\! \bs s $  \\
\hline

$B_{1u}$   & $\frac{\sigma_0\pm \sigma_z}{2}\otimes k_x\bs x \!\cdot\! \bs s - \frac{\sigma_0\mp \sigma_z}{2}\otimes k_y\bs y \!\cdot\! \bs s$,~~ $\sigma_x\otimes (k_x \bs y - k_y \bs x)\!\cdot\! \bs s $  \\
\hline

$B_{2u}$   & $\frac{\sigma_0\pm \sigma_z}{2}\otimes k_x\bs y \!\cdot\! \bs s+ \frac{\sigma_0\mp \sigma_z}{2}\otimes k_y\bs x \!\cdot\! \bs s$,~~ $\sigma_x\otimes (k_x \bs x + k_y \bs y)\!\cdot\! \bs s$  \\
\hline
\shortstack{ ~~ \\ $E_u$ \\ ~~ \\ ~~\\ ~~\\ ~~}        &

\shortstack{~~\\ $\left( i k_x \sigma_y,~~i k_y\sigma_y  \right)$\\
~~\\
$\left( \sigma_x\otimes k_y \bs{z}\!\cdot\!\bs s ,~~\sigma_x \otimes k_x\bs{z}\!\cdot\!\bs s \right)$  \\
~~\\
$ \left(\frac{\sigma_0 \pm \sigma_z}{2}\otimes k_x\bs{z}\!\cdot\!\bs s,~~\frac{\sigma_0 \mp \sigma_z}{2}\otimes  k_y\bs{z}\!\cdot\!\bs s \right)$ }
\\
\hline
\end{tabular}}
\label{tab:tab1}
\end{table*}

\section{Singlet-triplet-mixed even parity $A_{1g}$-pairing}
As one can see in Table \ref{tab:tab1}, there are multiple one-dimensional irreps which contain more than one symmetry-equivalent components and permit mixtures of spin-triplet and spin-singlet pairings. We take an example a simple $A_{1g}$ gap function,
\begin{equation}
\hat{\Delta}_{\bs k} = \psi_1 \hat{\Delta}_{1\bs k} + \psi_2 \hat{\Delta}_{2\bs k}  = ( \psi_1 \!\cdot\! i\sigma_y \otimes \bs z\!\cdot\! \bs s + \psi_2 \!\cdot\! \bs 1 ) is_y \,.
\label{eq:A1gGap}
\end{equation}
The triplet and singlet components correspond to inter- and intra-orbital pairings, respectively. In general, the two do not necessarily coexist in the absence of SOC -- when spins are good quantum numbers. To understand how SOC induces mixed pairings, we perform a standard free energy expansion, $f= \hat{\Delta}^\dagger\hat{\Delta}/V + T\sum_l \sum_{\bs k,w_n}\text{Tr}[G(iw_n,\bs k)\hat{\Delta}\bar{G}(iw_n,\bs k)\hat{\Delta}^\dagger]^{2l}/(2l)$ where $G(iw_n,\bs k)=(iw_n - H_{0\bs k})^{-1}$ and $\bar{G}(iw_n,\bs k)=(iw_n + H_{0,-\bs k}^\ast)^{-1}$ are the electron and hole components of the Gorkov Green's funcion. The singlet and triplet pairings are coupled at quadratic order,
\begin{equation}
J_{12}= i\lambda_{12} (\psi_1^\ast\psi_2 - \psi_2^\ast \psi_1)  \,,
\end{equation}
with $\lambda_{12} \propto \eta$ a real constant. The complex phase is a consequence of the particular structure of the SOC in Eq.~\ref{eq:H0}. A similar conclusion was reached in Ref.~\onlinecite{Puetter:12}. Therefore, SOC not only mixes but also selects a particular relative phase between the two components, e.g. $\theta_2-\theta_1= \pi/2$ if $\lambda_{12}>0$. The relative phase can be absorbed into the basis function. Thus a more compact form of Eq.~\ref{eq:A1gGap} reads: $\hat{\Delta}_{\bs k} \propto \left[ \sigma_y \otimes (\bs z\!\cdot\! \bs s) + \epsilon \bs 1  \right] i s_y$, where $\epsilon$ is a real constant determined by the details of the microscopic model. Notice there exists no ground state degeneracy, and such a pairing is time-reversal invariant (TRI), i.e. it satisfies $\mathcal{T}\hat{\Delta}_{\bs k} \mathcal{T}^{-1}= \hat{\Delta}_{-\bs k}$. On the contrary, the pairings with relative phases of $0$ and $\pi$ between $\psi_1$ and $\psi_2$ are degenerate and violate time reversal symmetry. It is also worth stressing that, in contrast to a pure spin-triplet state, a mixed singlet-triplet pairing may see a reduced uniform spin susceptibility below $T_c$. 

In like manner, the remaining two components of $A_{1g}$ given in Table \ref{tab:tab1}, $\hat{\Delta}_{3\bs k}= k_xk_y \sigma_x \otimes is_y$ and $\hat{\Delta}_{4\bs k}= (k_x^2-k_y^2)\sigma_z\otimes is_y$, also couple quadratically to the first two components, besides a coupling of similar order between themselves. In full, the free energy up to the quadratic order reads,
\begin{eqnarray}
f_\text{2nd}&=& \sum_{j=1}^4 \alpha_j|\psi_j|^2+i\sum_{j=2}^4 \left( \lambda_{1j} \psi_1^\ast \psi_j - c.c. \right) \nonumber \\
&+& \left(\lambda_{23}\psi_{2}^\ast\psi_3 + \lambda_{24} \psi_2^\ast\psi_4 + \lambda_{34} \psi_3^\ast\psi_4 + c.c \right)  \,.
\label{eq:A1gF2nd}
\end{eqnarray}
All of the $\alpha_j$ and $\lambda_{ij}$-coefficients are real. Like $\lambda_{12}$, the other two coefficients that couple triplet and singlet pairings, $\lambda_{13}$ and $\lambda_{14}$, both depend on SOC. By contrast, the remaining coefficients, $\lambda_{23},\lambda_{24}$ and $\lambda_{34}$, do not rely on SOC.  Instead, these three couplings are induced by the $\sigma_x$ and/or $\sigma_z$ terms in Eq. \ref{eq:H0}, with $\lambda_{23} \propto t^{\prime\prime}/t$, $\lambda_{24} \propto \tilde{t}/t$ and $\lambda_{34} \propto \tilde{t}t^{\prime\prime}/t^2$. The sign of $\alpha_i$ determines whether an intrinsic Cooper instability exists for the corresponding pairing component. The most negative $\alpha_i$ typically signifies the most dominant component. A component that lacks Cooper instability ($\alpha_i>0$) may still be induced due to the effective proximity effects through the finite couplings in the following sense. The free energy can be minimized by taking the lowest-energy eigenvalues of the coupling matrix, with the basis defined by $\hat{\psi}= (\psi_1,\psi_2,\psi_3,\psi_4)^T$:
\begin{equation}
f_\text{2nd} = \hat{\psi}^\dagger
\begin{bmatrix}
\alpha_1 & i\lambda_{12} & i\lambda_{13} & i \lambda_{14} \\
-i \lambda_{12} & \alpha_2 & \lambda_{23} & \lambda_{24} \\
-i \lambda_{13} & \lambda_{23} & \alpha_3 & \lambda_{34} \\
-i \lambda_{14} & \lambda_{24} & \lambda_{34} & \alpha_4
\end{bmatrix} \hat{\psi} \,.
\label{eq:A1gCouplingMatrix}
\end{equation}
In the single most favorable eigenstate, $\psi_1$ should acquire a relative phase of $\pi/2$ or $-\pi/2$ with respect to the remaining components. A general $A_{1g}$ gap function, with all of the four components emerging simultaneously, is then given by,
\begin{equation}
\hat{\Delta}_{\bs k} = i \epsilon_1 \hat{\Delta}_{1\bs k} + \epsilon_2 \hat{\Delta}_{2\bs k } +  \epsilon_3\hat{\Delta}_{3\bs k} + \epsilon_4 \hat{\Delta}_{4\bs k} \,,
\label{eq:A1gFullGap}
\end{equation}
where $(\epsilon_1, i\epsilon_2,i\epsilon_3,i\epsilon_4)$ constitutes the lowest-energy eigenvector of the coupling matrix in Eq.~\ref{eq:A1gCouplingMatrix}. In reality, one or certain subset of the $\epsilon_i$'s may dominate, while the rest are induced. For example, since $\hat{\Delta}_{2\bs k}$ and $\hat{\Delta}_{4\bs k}$ both describe intra-orbital pairing and since orbital-mixing is secondary to the intra-orbital hoppings in \SRO, $\epsilon_2$ and $\epsilon_4$ could be much larger than the others.

\section{spin-triplet even-parity $E_g$-pairing}
In single-orbital models, the ordinary $E_g$ pairing is even-parity and spin-singlet in nature, and it must involve out-of-plane pairing, taking the form of $k_z(k_x,k_y)$. However, in the present 2d two-orbital model, the simplest $E_g$ pairing taken from Table \ref{tab:tab1} is a spin-triplet given by,
\begin{equation}
\hat{\Delta}_{\bs k} =\left( \psi_{x}\!\cdot\! i\sigma_y\otimes \bs x\!\cdot\! \bs s + \psi_{y}\!\cdot\! i\sigma_y\otimes \bs y\!\cdot\! \bs s \right) is_y  \,,
\label{eq:Eg}
\end{equation}
where the two order parameters $\psi_x$ and $\psi_y$ form a two-dimensional irrep. This pairing has also been discussed in Ref.~\onlinecite{Cheung:19}. In essence, the two components each describes a spin-triplet inter-orbital $s$-wave pairing. A Ginzburg-Landau free energy can be constructed on symmetry basis or through a straightforward gradient expansion, which leads to,
\begin{eqnarray}
f&=& k_1 \left( \left|\partial_x \psi_x\right|^2+ \left|\partial_y \psi_y\right|^2 \right) + k_2 \left( \left|\partial_y \psi_x\right|^2+ \left|\partial_x \psi_y\right|^2 \right) \nonumber \\
&& +\alpha \left(|\psi_x|^2 + |\psi_y|^2 \right) + \beta \left( |\psi_x|^4+|\psi_y|^4 \right) \nonumber \\
&&+\beta_{xy}|\psi_x|^2|\psi_y|^2 + \beta^\prime \left[(\psi_x^\ast\psi_y)^2+ (\psi_y^\ast\psi_x)^2\right] +\cdots,~~~~~
\label{eq:fEg}
\end{eqnarray}
where `$\cdots$' denotes higher order terms. Note that because $\psi_{x}$ and $\psi_{y}$) are both even under spatial transformation $x \ra -x$ and $y\ra -y$ (or $k_x \ra -k_x$ and $k_y \ra -k_y$), cross-gradient terms such as $\partial_x\psi_x^\ast \partial_y\psi_y$  are disallowed. Likewise, $\partial_x\psi_x^\ast\partial_x\psi_y$, $\partial_y\psi_y^\ast\partial_y\psi_x$ and their complex conjugates are forbidden, as $\psi_x$ and $\psi_y$ exhibit opposite mirror eigenvalues about the $xz$ (and $yz$) planes. Dependent on the sign of $\beta^\prime$, two types of superconducting phases are possible, one preserving and the other breaking time reversal symmetry. When $\beta^\prime > 0$, the two components preferentially develop a relative phase of $\pm\pi/2$, leading to a TRSB pairing; whereas a relative phase of $0$ or $\pi$ is favored if $\beta^\prime<0$, which corresponds to a time-reversal invariant (TRI) state.

A TRSB pairing may support spontaneous current at the surface or around defects. Within Ginzburg-Landau theory, it is been well understood that the forbidden gradient terms mentioned above would have been crucial for the existence of spontaneous current \cite{Volovik:85,Sigrist:89,Furusaki:01,Huang:14}. Thus, unlike the conventional $E_g$ chiral d-wave pairing with $\Delta_{\bs k} \sim (k_x+ik_y)k_z$, the present TRSB $E_g$ pairing (when appears alone) has the salient feature that it is free of surface current. On the other hand, the system may exhibit superconducting domain walls separating regions of distinct TRSB pairings, and the neighboring corners of such domain walls carry opposite fractional quantum fluxes analogous to the scenario in a coupled anisotropic XY-model \cite{Isacsson:05,Xu:07}. The resultant internal field distribution could be detected in $\mu$SR measurements. Notably, fractional vortices could still emerge even when the pairing is TRI \cite{Huang:19}. A final important remark is that, since this spin-triplet pairing has its d-vector oriented in-plane, the Knight shift shall exhibit a drop under in-plane magnetic fields. This is potentially relevant to the observation in a recent NMR measurement~\cite{Luo:19}.

\section{singlet-triplet-mixed odd-parity $E_u$-pairing}
We write down in Table \ref{tab:tab1} four of the simplest basis functions belonging to the $E_u$ irrep. Note that, compared to the other terms, the second term, $\left( \sigma_x\otimes k_y \bs{z}\!\cdot\!\bs s ,~~\sigma_x \otimes k_x\bs{z}\!\cdot\!\bs s \right)$, has the form factors $k_x$ and $k_y$ in reversed order. In this manner, all four terms transform coherently as the basis $(k_y,k_x)$ does. Among the four terms, the first is the only spin-singlet pairing, and the third one is frequently discussed in connection to the proposal of p-wave instability on the quasi-1D bands \cite{Raghu:10b}. A general $E_u$-pairing acquires the form $\hat{\Delta}_{\bs k} = \sum_{i=1}^{4}\sum_{\mu=x,y}\psi_{i\mu} \hat{\Delta}_{i\bs k\mu}$. Following the analysis in the preceding section, we obtain the following free energy,
\begin{eqnarray}
f_\text{2nd} &=& \sum_{i=1}^4\sum_{\mu=x,y} \alpha_i |\psi_{i\mu}|^2 +\sum_{\mu=x,y} [ \lambda_{23} \psi_{2\mu}^\ast\psi_{3\mu} \nonumber \\
&&+ \lambda_{24} \psi_{2\mu}^\ast\psi_{4\mu}  + \lambda_{34} \psi_{3\mu}^\ast\psi_{4\mu}  +c.c.  ] \nonumber \\
&&+ i [ \psi_{1x}^\ast (\lambda_{12}\psi_{2x}+\lambda_{13}\psi_{3x}+ \lambda_{14}\psi_{4x}) - c.c.] \nonumber  \\
&&- i [ \psi_{1y}^\ast (\lambda_{12}\psi_{2y}+\lambda_{13}\psi_{3y}+ \lambda_{14}\psi_{4y}) - c.c.]
\label{eq:fEu}
\end{eqnarray}
where all $\lambda_{ij}$ are real quantities. In particular, $\lambda_{12},\lambda_{13}, \lambda_{14} \propto \eta$, demonstrating once again that SOC couples the singlet to the triplet pairings. The couplings between the triplet pairings (i.e. $\lambda_{23},\lambda_{24}, \lambda_{34}$) does not require finite SOC but needs other ingredients such as the inter-orbital hybridization $t^{\prime\prime}$. Finally, the last two lines indicate that the coupling between the singlet and triplet pairings has opposite signs for the $x$ and $y$-components. This has consequences on the phase configuration acquired by the multiple components. 

In short, denoting the the two corresponding order parameters $\Psi_{a(b)}$, the $E_u$ gap function is more generally expressed in an alternative two-component form: $\hat{\Delta}_{\bs k}=(\hat{\Delta}_{a\bs k},\hat{\Delta}_{b\bs k})$, with,
\begin{eqnarray}
\hat{\Delta}_{a\bs k} &=& \Psi_a ( i\epsilon_1 \hat{\Delta}_{1\bs k x} +  \epsilon_2 \hat{\Delta}_{2\bs k x} +  \epsilon_3\hat{\Delta}_{3\bs k x} +  \epsilon_4 \hat{\Delta}_{4\bs k x} ), \nonumber \\
\hat{\Delta}_{b\bs k} &=&\Psi_b ( -i\epsilon_1 \hat{\Delta}_{1\bs k y} +  \epsilon_2 \hat{\Delta}_{2\bs k y} +  \epsilon_3 \hat{\Delta}_{3\bs k y} + \epsilon_4 \hat{\Delta}_{4\bs k y} ), \nonumber \\
&&
\end{eqnarray}
where $\epsilon_{1,\cdots,4}$ are real constants. This leads to the following free energy in powers of $\Psi_a$ and $\Psi_b$,
\begin{eqnarray}
f &=& k_1 \left(\left|\partial_x\Psi_a\right|^2 + \left|\partial_y\Psi_b\right|^2 \right) + k_2 \left(\left|\partial_x\Psi_b\right|^2 + \left|\partial_y\Psi_a\right|^2 \right)  \nonumber \\
&&+ k_3\left(\partial_x\Psi_a^\ast \partial_y\Psi_b  + c.c. \right) + k_4\left(\partial_x\Psi_b^\ast \partial_y\Psi_a  + c.c. \right)  \nonumber \\
&&+\alpha \left(|\Psi_a|^2 + |\Psi_b|^2 \right) + \beta\left(|\Psi_a|^4 + |\Psi_b|^4 \right) \nonumber \\
&& + \beta_{ab}|\Psi_a|^2|\Psi_b|^2 + \beta^\prime\left[ (\Psi_a^\ast \Psi_b)^2 + (\Psi_b^\ast \Psi_a)^2  \right] \!+\! \cdots. \nonumber \\
&&   
\label{eq:fEu2}
\end{eqnarray}
Compared to the effective theory in Eq. \ref{eq:fEg}, the cross-gradient terms with coefficients $k_3$ and $k_4$ are present, and they could generate finite spontaneous current if the pairing breaks time-reversal symmetry. On the other hand, the singlet-triplet mixing shall lead to a suppressed uniform spin susceptibility and therefore a drop in NMR Knight shift under in-plane fields. Further, Ref.~\onlinecite{Li:19} will study some peculiar forms of the multi-orbital $E_u$ pairing, which exhibits near-nodal excitations consistent with a number of experimental signatures~\cite{Nishizaki:00,Hassinger:17}.  

\section{Three-orbital model}
\label{sec:3orbital}
In extending to a full three-orbital model, the Gell-Mann matrices ($T_i$, $i=1,...,8$) turn out to be convenient devices. We define $\bar{T}_{11}=(T_0+\sqrt{3}T_8)/2$ and $\bar{T}_{33}=(T_0-\sqrt{3}T_8)/4$, where $T_0$ is a $3\times 3$ identity matrix. Using the orbital spinor basis $(c^\dagger_{m\ua},c^{\dagger}_{m\da})$ in the order $m=xz,yz,xy$, up to quadratic order in $k$ and with on-site SOC, the Hamiltonian reads,
\begin{eqnarray}
H_{0\bs k}&=&\left[ t(k_x^2+k_y^2)-\mu \right] \bar{T}_{11}+ \tilde{t}(k_x^2-k_y^2)T_3 + t^{\prime\prime}k_xk_y T_1  \nonumber \\
 &+& \left[ t^\prime (k_x^2 + k_y^2)-\mu_{xy}  \right]\bar{T}_{33} \nonumber \\
&+&  \eta(T_2\otimes s_z + T_5\otimes s_x - T_7 \otimes s_y ) \,,
\label{eq:H3}
\end{eqnarray}
where the $t^\prime$ term and $\mu_{xy}$ denote the kinetic energy and chemical potential of the $d_{xy}$-orbital. Note that $T_{1,2,3}$ are equivalent to $\sigma_{x,y,z}$, and $\bar{T}_{11}$ to $\sigma_0$. Hence they inherit the transformation properties of the $\sigma_\mu$-operators. $T_{4,5}$ and $T_{6,7}$, on the other hand, transform respectively as the $B_{3g}$ and $B_{2g}$ irreps of the $D_{2h}$ group. However, the SOC term (the last term), having an appropriate linear superposition of $T_5$ and $T_7$, respects $D_{4h}$.

Without further elaboration, the gap functions, especially those not involving inter-orbital pairings with the $d_{xy}$ orbital, can be classified rather straightforwardly following the preceding analyses. Inter-orbital pairings involving $d_{xy}$ are associated with pairing operators $T_{4,...,7}$. As can be checked, $(T_{4},T_6)$ [and $(T_5,T_7)$] transform as $E_g$ ($E_u$) irrep under $D_{4h}$. As a consequence, any such pairing must contain both $T_{4}$ and $T_6$ (or $T_5$ and $T_7$) in the gap function. This is demonstrated in Table \ref{tab:tab3}.  As an interesting note, two recent microscopic multi-orbital calculations \cite{Wang:18,Gingras:18} both found noticeable, or even dominant, interorbital $E_g$ or $E_u$ pairing involving the $d_{xy}$-orbital in some regimes of the interaction parameter space.

\begin{table}
\caption{\label{tab:table1}  Representative superconducting basis functions of the inter-orbital pairing involving the $d_{xy}$-orbital in the two dimensional $E_g$ and $E_u$ irreps. Here $T_i$ and $s_i$ operate respectively in the orbital and spin space, as explained in the text. Note that in the $E_g$ irrep, the order of the two degenerate components is to ensure that each basis function transforms as $(k_y,k_x)k_z$ does, instead of some behaving like $(k_y,k_x)k_z$ and some like  $(k_x,k_y)k_z$. Similarly for the $E_u$ basis functions. }
{\renewcommand{\arraystretch}{1.3}
\begin{tabular}{c|c}
\hline
~~~~irrep.~~~~ & ~~~~~~~~~basis function ~~~~~~~\\
\hline
$A_{1g}$   & $ T_5\otimes \bs x\!\cdot\! \bs s - T_7\otimes \bs y\!\cdot\! \bs s $ \\
\hline

$A_{2g}$   & $ T_5\otimes \bs y\!\cdot\! \bs s +T_7\otimes \bs x\!\cdot\! \bs s$  \\
\hline

$B_{1g}$   &   $T_5\otimes \bs x\!\cdot\! \bs s + T_7\otimes \bs y\!\cdot\! \bs s$  \\
\hline

$B_{2g}$   &   $T_5\otimes \bs y\!\cdot\! \bs s - T_7\otimes \bs x\!\cdot\! \bs s $  \\
\hline
\shortstack{ ~~ \\ $E_g$ \\ ~~ \\ ~~\\~~\\~~}             &

\shortstack{~~ \\ $ \left( T_4,~~ T_6 \right)$ \\
~~\\
$ \left[ k^2_{x(y)} T_4,~~ k^2_{y(x)}T_6 \right]  $ \\
~~\\
$\left(iT_7\otimes \bs z \!\cdot\! \bs s, iT_5\otimes \bs z \!\cdot\! \bs s \right) $  \\
~~\\
$\left[ i T_7\otimes k^2_{x(y)} \bs z \!\cdot\! \bs s, iT_5\otimes k^2_{y(x)}\bs z \!\cdot\! \bs s \right] $ }  \\
\hline

\shortstack{ ~~ \\ $E_u$ \\ ~~ \\ ~~\\ ~~\\ ~~}        &

\shortstack{~~\\ $\left( T_4\otimes k_x \bs x\!\cdot\! \bs s,~~T_6\otimes k_y \bs y\!\cdot\! \bs s \right)$ \\
~~\\
$ \left(T_6\otimes k_y \bs x\!\cdot\! \bs s,~~T_4\otimes k_x \bs y\!\cdot\! \bs s  \right)$ \\
~~\\
$\left( T_4\otimes k_y \bs y\!\cdot\! \bs s,~~T_6\otimes k_x \bs x\!\cdot\! \bs s  \right)$ \\
~~\\
$\left( T_6\otimes k_x \bs y\!\cdot\! \bs s,~~T_4\otimes k_y \bs x\!\cdot\! \bs s \right)$ } \\
\hline
\end{tabular}}
\label{tab:tab3}
\end{table}

\section{Summary and discussions}
\label{sec:summary}
With an eye on the yet-unresolved myth of the superconducting \SRO, we explored the possibilities made available by its multi-orbital degree of freedom. The superconducting pairings are classified on the basis of the Ru $t_{2g}$ 4$d$ orbitals according to the underlying crystal point group symmetries. This leads to multiple exotic superconducting pairings not accessible in single-orbital or itinerant-electron models. In some cases, the phenomenology of the orbital-basis description could differ considerably from that of an itinerant-band description. We discussed some of their salient aspects and made connections to \SRO~in due course. As a special note, when a spin-triplet pairing is inherently mixed with spin-singlet pairings due to the presence of SOC, the uniform spin susceptibility, and hence the Knight shift, may exhibit a drop below the superconducting transition. 

Our main purpose is not to rule out or identify any pairing for \SRO, but rather to provide a new perspective to further explore the enigmatic superconductivity in this material. Hence we have restricted, for simplicity, to in-plane pairings in our symmetry classification of the multi-orbital superconductivity. Including out-of-plane couplings, i.e. extending the model to three spatial dimensions (3d), brings about numerous additional possibilities. In fact, even within the conventional band description, some novel forms of pairings may arise due to a 3d spin-orbital entanglement in the electronic structure. In particular, the $E_u$ pairing is recently shown to be inherently three-dimensional \cite{Huang:18,Huang:19}, containing both in-plane and out-of-plane pairings. This is unlike what has been typically assumed for quasi-2d models. More intriguingly, a 3d nematic $E_u$ pairing, which can be realized if the out-of-plane pairing is sizable, was argued to explain a number of outstanding puzzles, such as the absence of surface current and the anomalous response to in-plane uniaxial strains \cite{Huang:18,Huang:19}. Notably, since the d-vector of a 3d $E_u$ pairing has both in-plane and out-of-plane components, a drop in the NMR Knight shift is expected for generic in-plane magnetic field orientations \cite{Huang:19}. Additionally, models containing out-of-plane pairings have appeared in several other contexts \cite{Hasegawa:00,Zhitomirsky:01,Annett:02}.

{\bf Note added-} As this manuscript was being prepared for submission, a preprint appeared on arXiv\cite{Ramires:19} with a similar idea to exploit the multi-orbital nature of the superconductivity in \SRO. Later, another similar paper also appeared on arXiv \cite{Kaba:19}.  

\acknowledgements
We would like to thank Fu-Chun Zhang and Qiang-Hua Wang for valuable discussions. This work is supported in part by the NSFC under grants No. 11825404 (W.H. and H.Y.), No. 11774306 and No. 11704106 (Y.Z.), the National Key Research and Development Program of China under grant No.2016YFA0300202 (Y.Z.) and No. 2016YFA0301001 (H.Y.), the Strategic Priority Research Program of Chinese Academy of Sciences under Grant No. XDB28000000 (Y.Z. and H.Y.), a startup grant at SUSTech and the C.N. Yang Junior Fellowship at Tsinghua University (W.H.).

\end{document}